\documentclass[12pt]{iopart}

\usepackage{graphicx}
\begin{document}

\title{Orbital Kondo effect in a parallel double quantum dot}

\author{Zhi-qiang Bao$^1$, Ai-Min Guo$^1$, and Qing-feng Sun$^{2,3}$}
\address{$^1$ Institute of Physics, Chinese Academy of Sciences, Beijing 100190, China}
\address{$^2$ International Center for Quantum Materials, School of Physics, Peking University, Beijing 100871, China}
\address{$^3$ Collaborative Innovation Center of Quantum Matter, Beijing 100871, China}
\ead{sunqf@pku.edu.cn}

\begin{abstract}
We construct a theoretical model to study the orbital Kondo effect
in a parallel double quantum dot (DQD). Recently,
pseudospin-resolved transport spectroscopy of the orbital Kondo
effect in a DQD has been experimentally reported. The experiment
revealed that when interdot tunneling is ignored, there exist two
and one Kondo peaks in the conductance-bias curve for the
pseudospin-non-resolved and pseudospin-resolved cases, respectively.
Our theoretical studies reproduce this experimental result. We also
investigate the situation of all lead voltages being non-equal (the
complete pseudospin-resolved case), and find that there are four
Kondo peaks at most in the curve of the conductance versus the
pseudospin splitting energy. When the interdot tunneling is
introduced, some new Kondo peaks and dips can emerge. Besides, the
pseudospin transport and the pseudospin flipping current are also
studied in the DQD system. Since the pseudospin transport is much
easier to be controlled and measured than the real spin transport,
it can be used to study the physical phenomenon related to the spin
transport.
\end{abstract}

\maketitle

\section{Introduction}
The Kondo effect is an important issue in condensed-matter
physics \cite{book1} and has been attracted extensive attention since
its first discovery, because the Kondo effect could provide a deeper
understanding of the physical properties of many strong correlated
systems \cite{pw14-33}. On the other hand, a quasi-zero-dimensional
system called quantum dot (QD), of which the parameters can be
modulated experimentally in a continuous and reproducible manner,
offers proper platform to study the Kondo
problems \cite{science281-526,science289-2105,nature391-156,science281-540,nature405-764}.
Under appropriate conditions, the Kondo effect can arise from the
coherent superposition of the cotunneling
processes \cite{pw14-33,science281-540}, where the spin degree of
freedom plays a significant role and the electron in the QD can flip
its spin. At low temperature, the coherent superposition of many
cotunneling processes could lead to the Kondo resonant state in
which the spin flip occurs frequently within the QD and a very sharp
Kondo peak emerges in the density of state of the QD.

Later on, the Kondo effect was proposed based on the orbital degree
of freedom \cite{prb66-155308,
Physica14E-385,prb70-075204,nature434-484,prb74-233301,prl99-066801,prl110-046604}.
It was reported that double QD (DQD) could become a good candidate
for realizing the orbital Kondo effect \cite{prb66-155308,prl110-046604,prb88-235427,prb88-245130,np10-145,prb85-241310,apl104-132401,prl101-186804,prb84-161305}.
In this situation, the energy of the orbital state in the left QD
can be the same as or very close to that in the right QD. Then, the
corresponding left and right orbital states are degenerate or near
degenerate, and they can be regarded as pseudospin degenerate
states \cite{prb66-155308,physica21E-1046,prl105-246804}. In real
spin systems, it is difficult to manipulate the spin-up state and
the spin-down one individually. In contrast, since the left and
right QDs of the DQD system are separated in space, it is much
easier to control over both QDs and each of them can be seen as a
pseudospin
component \cite{prl110-046604,prb85-241310,apl85-1846,prl106-106401,prb71-115312,prl90-026602}.
As a result, the physical phenomenon, which is related to the spin
degree of freedom, may also be realized in the DQD system including
the pseudospin (orbital) degree of freedom.

\begin{figure}
\centering
\includegraphics[width=0.6\textwidth]{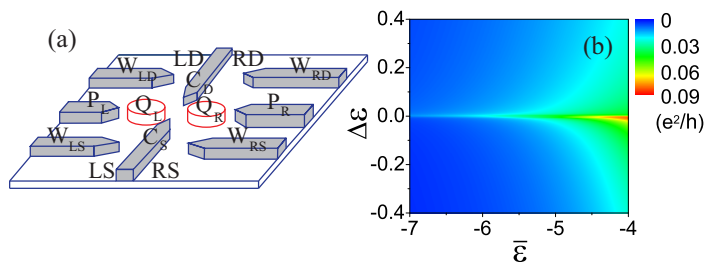}
\caption{\label{fig1} (a) Schematic diagram for a parallel DQD
device. Q$_L$ and Q$_R$ are the left and right QDs. The gates
P$_{L}$ and P$_{R}$ are used to control the occupancy of the QDs.
The tunneling rates between dot Q$_{L}$ (Q$_{R}$) and its source
lead $LS$ ($RS$) and drain lead $LD$ ($RD$) are controlled by gates
W$_{LS}$ (W$_{RS}$) and W$_{LD}$ (W$_{RD}$), respectively. The gates
C$_{S}$ and C$_{D}$ are used to control the interdot tunneling. (b)
$G_{LS}$ as functions of $\bar{\varepsilon}$ and
$\Delta\varepsilon$. The source and drain voltages are
$V_{LS}=V_{RS}=V_{LD}=V_{RD}=0$. Here, the temperature keeps
$T=0.001$ and $t_{c}=0$.}
\end{figure}

Very recently, the pseudospin-resolved transport spectroscopy of the
Kondo effect has been observed in a DQD device on the basis of a
orbital degeneracy \cite{prl110-046604}. The schematic diagram of
this device is shown in figure \ref{fig1}(a). In the experiment, the
authors fabricated the parallel DQD system from an epitaxially grown
AlGaAs/GaAs heterostructure. As illustrated in figure \ref{fig1}(a),
Q$_{L}$ and Q$_{R}$ are the parallel QDs, which are capacitively
coupled with each other. The voltages applied on the gates P$_{L}$
and P$_{R}$ are used to control the occupancy of the dots. The gates
W$_{LS}$ (W$_{RS}$) and W$_{LD}$ (W$_{RD}$) control the tunneling
rates between dot Q$_{L}$ (Q$_{R}$) and its source lead $LS$ ($RS$)
and drain lead $LD$ ($RD$). The gates C$_{S}$ and C$_{D}$ are used
to control the interdot tunneling. In \cite{prl110-046604}, the authors applied negative
voltages on the gates C$_{S}$ and C$_{D}$ to make the interdot
tunneling negligible. They measured the standard transport
spectroscopy as a function of the bias voltages and observed a
zero-bias peak in the conductance. Furthermore, if the orbital
degeneracy is broken, the Kondo resonances have different pseudospin
character. Using pseudospin-resolved spectroscopy, they observed a
Kondo peak at only one sign of the bias voltage.

In this paper, we theoretically investigate the orbital Kondo effect
in a parallel DQD.
It need to mention that the properties of DQDs
have been studied by lots of theoretical works \cite{prb88-235427,prb88-245130,np10-145,prb85-241310,apl104-132401}.
In this work, by using the non-equilibrium Green's function
method and the equation of motion technique, the formula of the
conductance for each pseudospin component and the pseudospin
flipping current are obtained.
The main new results are listed as follows: (1) If the interdot tunneling
coupling $t_{c}$ is zero, we reproduce the experimental results in
\cite{prl110-046604}, where two Kondo peaks were observed in the
conductance-bias curve for the pseudospin-non-resolved case and
only one Kondo peak was found in the pseudospin-resolved case. In
the curve of the conductance versus the pseudospin splitting energy,
there exist three and two Kondo peaks for the pseudospin-non-resolved
and pseudospin-resolved cases, respectively. (2) When the interdot
tunneling coupling $t_{c}$ is nonzero, the levels in the DQD can
form molecular states. Then, the Kondo peaks can emerge at
$\Delta E=\pm V_{LS}$ for both pseudospin-non-resolved and
pseudospin-resolved cases, where $\Delta E$ is the energy difference
between the two molecular states and $V_{LS}$ is the lead voltage.
Besides, an additional Kondo peak and dip structure could emerge at
$\Delta E=0$. (3) The pseudospin transport and the pseudospin
flipping current in the DQD system are studied. In particular, the
pseudospin system is much easier to be controlled and measured than
the real spin current, so the pseudospin DQD system can be a good
candidate for studying the properties related to the spin degree of
freedom.

The rest of the paper is organized as follows. In section
\ref{sec:models}, we propose the model Hamiltonian, and use the
non-equilibrium Green's function method to get the current and
conductance formulas. In section \ref{sec:discussions}, we numerically
investigate the conductances and the pseudospin flipping current of
the DQD in different cases. Finally, we give the conclusions in section
\ref{sec:conclusions}.

\subsection{Model and analytical results}
\label{sec:models}

The Hamiltonian of the DQD system as shown in figure \ref{fig1}(a) can be
written as
\begin{eqnarray}
H=H_{DQD}+H_{T}+\sum\limits_{\alpha\beta}H_{\alpha\beta},\label{eq:1a}
\end{eqnarray}

where
\begin{eqnarray}
H_{DQD}&=& \sum\limits_{\alpha}\varepsilon_{\alpha}d^ {\dagger}_ {\alpha}d_{\alpha}+ Ud^{\dagger}_{L}d_{L}d^ {\dagger}_{R} d_{R}+(t_{c}d^{\dagger}_{L}d_{R}+h.c.), \nonumber\\ H_{T}&=&\sum\limits_{\alpha\beta k}(t_{\alpha\beta}a^{\dagger}_ {\alpha\beta k}d_{\alpha}+h.c.), \nonumber\\ H_{\alpha\beta} &=&\sum\limits_{k} \varepsilon_{\alpha\beta k}a^{\dagger}_{\alpha\beta k}a_{\alpha\beta k}. \nonumber
\end{eqnarray}
Here, $H_{DQD}$ is the Hamiltonian of the DQD and $d_\alpha^
\dagger$ ($d_\alpha$) is the creation (annihilation) operator of the
electron in the QDs with $\alpha=L/R$ representing left and right.
$\varepsilon_\alpha$ is the energy level of the QDs, $U$ is the
interdot electron-electron interaction, and $t_c$ is the tunneling
coupling. $H_{T}$ denotes the tunneling between the DQD and the
leads. $a^{\dagger}_{\alpha\beta k}$($a_{\alpha\beta k}$) is the
creation (annihilation) operator of the electron in the leads, with
$\beta=S/D$ being source and drain. $H_{\alpha\beta}$ describes the
noninteracting leads. It should be noted that when a high magnetic
field is applied to the QD, the spin splitting energy can be
comparable with or even larger than the QD energy level
spacing \cite{prb66-155308,prl82-2931,science278-1788}. Then, as
compared with the low energy spin state, the opposite spin state
does not affect the transport property of the system at low bias.
Thus, we can neglect the spin degree of freedom. Meanwhile, we
consider the low bias case with its value being less than the
intradot electron-electron interaction energy $U_0$. In this case,
there is only one eigenstate in each QD in the bias window. Then, we
can absorb the intradot interaction $U_{0}$ into the energy levels
$\varepsilon_{L}$ and
$\varepsilon_{R}$ \cite{prb66-155308,prl76-1916}. As a result, both
the spin degree of freedom and the intradot interaction can be
ignored, leaving only the interdot interaction $U$. This
approximation has been adopted in \cite{prb66-155308,prl76-1916}.

This model can describe various properties of the parallel DQD,
including the properties illustrated in the recent experiment of
\cite{prl110-046604}, and could also be used to study
the pseudospin transport. Here, we briefly introduce the concept of
the pseudospin transport in the parallel DQD. As we know, the
electron has two spin states: spin up and spin down. The transport
related to spin is called spin transport. Similarly, the electron in
the DQD also has two states: the electron in the left QD and that in
the right QD. When the electron is in the left (right) QD, we can
call it the pseudospin up (down) state. The transport related to the
pseudospin is called the pseudospin transport. There are four
advantages of the pseudospin transport: (1) The electrons with
different spins have the same chemical potential in the wires, i.e.,
$\mu_{\uparrow}=\mu_{\downarrow}$. Thus, it is difficult to
manipulate the electrons with specific spin while keeping the other
spin unchanged. Even if we can achieve the case of
$\mu_{\uparrow}\neq\mu_{\downarrow}$ by some special
methods \cite{suna1,suna2}, the spin voltage
$\mu_{\uparrow}-\mu_{\downarrow}$ is still difficult to control.
While regarding the pseudospin, however, the situation is totally
different. The voltages applied on the wires connected to the left
and right QDs can be manipulated separately. This means that the
chemical potentials of the electrons with different pseudospins,
$\mu_{L}$ and $\mu_{R}$, can be easily controlled, which has been
realized in the experiment \cite{prl110-046604}. (2) Since the energy
levels of different QDs can be manipulated by the gates P$_{L}$ and
P$_{R}$, the pseudospin splitting energy
$\Delta\varepsilon=\varepsilon_{L}-\varepsilon_{R}$ can be adjusted
in a wide range. (3) The pseudospin flipping strength $t_c$ in the
DQD can also be tuned by the gates C$_D$ and C$_S$. It can be open
or closed by simply tuning the gate voltages. (4) The real spin in
the lead is difficult to keep its direction and the spin flipping
exists inevitably. Contrarily, the pseudospin can keep its
``direction" steadily outside the DQD because the electrons in the
left lead cannot tunnel into the right lead and vice versa. So the
pseudospin flipping current in the DQD can accurately be measured in
the experiment.

Next, we will use the standard equation of motion technique to solve
the retarded Green's function \cite{prb66-155308,add1,suna3,suna4}.
The equation of motion is:
\begin{eqnarray}
\varepsilon\langle\langle A\vert B\rangle\rangle^{r}=\langle\{\hat{A}, \hat{B}\}\rangle+\langle\langle[\hat{A},H]\vert\hat{B} \rangle\rangle^{r},\label{eq:1b}
\end{eqnarray}
where $\hat{A}$ and $\hat{B}$ are arbitrary operators, and
$\langle\langle A\vert B\rangle\rangle^{r}$ is the standard notation
of the retarded Green's function. Since higher order Green's functions
will appear in the calculations of equation of motion, a decoupling
schemes is needed. The decoupling scheme in this work takes the following
rules: (1) if we use $X$ to represent the leads operator ($a_{\alpha\beta k}$
and $a^{\dagger}_{\alpha\beta k}$) and use $Y$ to represent the DQDs
operator ($d_{\alpha}$ and $d^{\dagger}_{\alpha}$), then we take $\langle XY\rangle=0$.
(2) If the two-particle Green's function involves two leads operators,
then we take $\langle\langle X_{1}X_{2}Y|d^{\dagger}_{\alpha}\rangle\rangle^{r}=\langle X_{1}X_{2}\rangle\langle\langle Y|d^{\dagger}_{\alpha}\rangle\rangle^{r}$.
(3) If the two-particle Green's function involves only one leads
operator, which is $\langle\langle XY_{1}Y_{2}|d^{\dagger}_{\alpha}\rangle\rangle^{r}$,
we continue to apply the equation of motion until all the two-particle
Green's functions contain two leads operators. This decoupling
scheme has been used in previous papers \cite{prb66-155308,add1}.

Moreover, because the method of derivation we used are similar to
Ref. \cite{prb66-155308}, we omit the detailed derivation and only show
the results in this paper. It should be pointed out that although
we use the same calculation method with Ref. \cite{prb66-155308},
the research subject and conclusions are totally different.
Ref. \cite{prb66-155308} described the series DQDs, while the present
work refers to the parallel DQDs.
Unlike the series DQDs in Ref. \cite{prb66-155308},
the conductance can hold a pseudospin-resolved
character in the paralell DQDs system when the interdot tunnelling $t_{c}=0$.
The pseudospin transport and the pseudospin flipping current
can also be studied in the present model while $t_{c}\neq 0$, and
the calculation presents a new method to measure and
control the pseudospin transport in the parallel DQDs system.

In addtion, it is worth mentioning that although the equation of
motion method based on non-equilibrium Green's function cannot
quantitatively obtain the intensity of the Kondo effect, it can give
the qualitative physics and the positions of the Kondo peaks. Using
the equation of motion in equation (\ref{eq:1b}), we can obtain the
matrix equation:
\begin{eqnarray}
&\left( \begin{array}{cc}
C_{11} &  C_{12}\\
C_{21} & C_{22} \\
\end{array} \right)
\cdot\left( \begin{array}{cc}
\langle\langle d_{L}\vert d_{L}^{\dagger}\rangle\rangle^{r} & \langle\langle d_{L}\vert d_{R}^{\dagger}\rangle\rangle^{r} \\
\langle\langle d_{R}\vert d_{L}^{\dagger}\rangle\rangle^{r} & \langle\langle d_{R}\vert d_{R}^{\dagger}\rangle\rangle^{r} \\
\end{array} \right)
=\left( \begin{array}{cc}
D_{11} & D_{12} \\
D_{21} & D_{22} \\
\end{array} \right),\label{eq:1g}
\end{eqnarray}
where
\begin{eqnarray}
C_{11}=&\varepsilon-\varepsilon_{L}-\Sigma_{LS}^{0}-\Sigma_{LD}^{0}+UA_{L}B
  \left(\tilde{t}_{c}A_{R}\Sigma^{d} \right.\nonumber \\ &  \left. + \Sigma_{RS}^{a}+\Sigma_{RD}^{a}+ \Sigma_{RS}^{b}+ \Sigma_{RD}^{b}+ \Sigma_{LS}^{c}+\Sigma_{LD}^{c}\right), \nonumber\\
C_{12}=&-t_{c}+UA_{L}B\left[\tilde{t}_{c}A_{R}\left(\Sigma_{LS}^{a}+ \Sigma_{LD}^{a}+\Sigma_{LS}^{b} \right.\right.\nonumber \\
&\left.\left. +\Sigma_{LD}^{b}+\Sigma_{RS}^{c}+\Sigma_{RD}^{c}\right)+ \Sigma^{d}\right], \nonumber\\
C_{21}=&-t_{c}+UA_{R}B\left[\tilde{t}_{c}A_{L}\left(\Sigma_{RS}^{a} +\Sigma_{RD}^{a}+\Sigma_{RS}^{b} \right.\right.\nonumber \\
&\left.\left.+\Sigma_{RD}^{b}+\Sigma_{LS}^{c}+\Sigma_{LD}^{c}\right) +\Sigma^{d}\right], \nonumber\\
C_{22}=&\varepsilon-\varepsilon_{R}-\Sigma_{RS}^{0}-\Sigma_{RD}^{0}+UA_{R}B \left(\tilde{t}_{c}A_{L}\Sigma^{d} \right.\nonumber \\
&\left.+\Sigma_{LS}^{a}+\Sigma_{LD}^{a}+\Sigma_{LS}^{b}+ \Sigma_{LD}^{b}+\Sigma_{RS}^{c}+\Sigma_{RD}^{c}\right), \nonumber\\
D_{11}=&1+UA_{L}Bn_{R}-UA_{L}B\tilde{t}_{c}A_{R}\langle d_{L}^{\dagger}d_{R}\rangle, \nonumber\\
D_{12}=&-UA_{L}B\langle d_{R}^{\dagger}d_{L}\rangle+UA_{L}B\tilde{t}_{c}A_{R}n_{L}, \nonumber\\
D_{21}=&-UA_{R}B\langle d_{L}^{\dagger}d_{R}\rangle+UA_{R}B\tilde{t}_{c}A_{L}n_{R}, \nonumber\\
D_{22}=&1+UA_{R}Bn_{L}-UA_{R}B\tilde{t}_{c}A_{L}\langle d_{R}^{\dagger}d_{L}\rangle. \nonumber
\end{eqnarray}
The expressions of the above notations are listed as follows:
\begin{eqnarray}
\Sigma_{\alpha\beta}^{0}&=&\sum_{k}\frac{\vert t_{\alpha\beta}\vert^2}{\varepsilon-\varepsilon_{\alpha\beta k}} = -\frac{i}{2}\Gamma_{\alpha\beta}, \nonumber\\
\tilde{\varepsilon}_{\alpha\beta k}^2&=&(\varepsilon+\varepsilon_{L}-\varepsilon_{R}-\varepsilon_{\alpha\beta k})(\varepsilon-\varepsilon_{L}+\varepsilon_{R}-\varepsilon_{\alpha\beta k})-4t_{c}^2, \nonumber\\
\Sigma_{\alpha\beta}^{1/a} &= &\sum_{k}\frac{\vert t_{\alpha\beta}\vert^2}{\varepsilon-\varepsilon_{L}-\varepsilon_{R} +\varepsilon_{\alpha\beta k}-U}F_{\alpha\beta}^{1/a}(\varepsilon_{\alpha\beta k}), \nonumber\\
\Sigma_{\alpha\beta}^{2/b}&=&\sum_{k}\frac{(\varepsilon- \varepsilon_{\alpha\beta k})(\varepsilon-\varepsilon_{\alpha}+\varepsilon_{\bar{\alpha}}- \varepsilon_{\alpha\beta k})-2t_{c}^2}{(\varepsilon-\varepsilon_{\alpha\beta k})\tilde{\varepsilon}_{\alpha\beta k}^2}\cdot\vert t_{\alpha\beta}\vert^{2}F_{\alpha\beta}^{1/a}(\varepsilon_{\alpha\beta k}), \nonumber\\
\Sigma_{\alpha\beta}^{3/c}&=&\sum_{k}\frac{2t_{c}^2}{(\varepsilon- \varepsilon_{\alpha\beta k})\tilde{\varepsilon}_{\alpha\beta k}^2}\vert t_{\alpha\beta}\vert^2F_{\alpha\beta}^{1/a}(\varepsilon_{\alpha\beta k}), \nonumber\\
\Sigma_{\alpha\beta}^{4/d}&=&\sum_{k}\frac{\varepsilon- \varepsilon_{\alpha}+\varepsilon_{\bar{\alpha}}-\varepsilon_{\alpha\beta k}}{(\varepsilon-\varepsilon_{\alpha\beta k})\tilde{\varepsilon}_{\alpha\beta k}^2}t_{c}\vert t_{\alpha\beta}\vert^2F_{\alpha\beta}^{1/a}(\varepsilon_{\alpha\beta k}), \nonumber\\
A_{\alpha}^{-1}&=&\varepsilon-\varepsilon_{\alpha}-U-\Sigma_{\alpha S}^{0}-\Sigma_{\alpha D}^{0}-\Sigma_{\bar{\alpha}S}^{1}-\Sigma_{\bar{\alpha}D}^{1} \nonumber \\
&&-\Sigma_{\bar{\alpha}S}^2-\Sigma_{\bar{\alpha}D}^{2}-\Sigma_{\alpha S}^{3}-\Sigma_{\alpha D}^{3}, \nonumber\\
\Sigma^{d}&=&\Sigma_{LS}^{d}+\Sigma_{LD}^{d}+\Sigma_{RS}^{d}+ \Sigma_{RD}^{d}, \nonumber\\
\tilde{t}_{c}&=&t_{c}+\Sigma_{LS}^{4}+\Sigma_{LD}^{4}+ \Sigma_{RS}^{4}+\Sigma_{RD}^{4}, \nonumber\\
B^{-1}&=&1-\tilde{t}_{c}^{2}A_{L}A_{R}. \nonumber
\end{eqnarray}
In the above equations, $F_{\alpha\beta}^{1} (\varepsilon_
{\alpha\beta k})=1$ and
$F_{\alpha\beta}^{a}(\varepsilon_{\alpha\beta
k})=f_{\alpha\beta}(\varepsilon_{\alpha\beta k})$, where
$f_{\alpha\beta}(\varepsilon_{\alpha\beta
k})=1/\{\exp[(\varepsilon_{\alpha\beta
k}-\mu_{\alpha\beta})/k_BT]+1\}$ is the Fermi distribution function
and $\mu_{\alpha\beta}=eV_{\alpha\beta}$ is the chemical potential
of the lead $\alpha\beta$. $\alpha$ and $\bar{\alpha}$ denote
different left-right positions. That is, if $\alpha$ is left,
$\bar{\alpha}$ is right; if $\alpha$ is right, $\bar{\alpha}$ is
left. $\Sigma_{\alpha\beta}^{1}$, $\Sigma_{\alpha\beta}^{2}$,
$\Sigma_{\alpha\beta}^{3}$, $\Sigma_{\alpha\beta}^{4}$,
$\Sigma_{\alpha\beta}^{a}$, $\Sigma_{\alpha\beta}^{b}$,
$\Sigma_{\alpha\beta}^{c}$, and $\Sigma_{\alpha\beta}^{d}$ are the
higher-order self-energies.

Taking the limit of $U\rightarrow\infty$, equation (\ref{eq:1g}) can be
simplified and the elements of the matrices are replaced by:
\begin{eqnarray}
C_{11}&=&\varepsilon-\varepsilon_{L}-\Sigma_{LS}^{0}-\Sigma_{LD}^{0}-\Sigma_{RS}^{b}-\Sigma_{RD}^{b}-\Sigma_{LS}^{c}-\Sigma_{LD}^{c}, \nonumber\\
C_{12}&=&-t_{c}-\Sigma^{d}, \nonumber\\
C_{21}&=&-t_{c}-\Sigma^{d}, \nonumber\\
C_{22}&=&\varepsilon-\varepsilon_{R}-\Sigma_{RS}^{0}-\Sigma_{RD}^{0}-\Sigma_{LS}^{b}-\Sigma_{LD}^{b}-\Sigma_{RS}^{c}-\Sigma_{RD}^{c}, \nonumber
\end{eqnarray}
and
$D_{11}=1-n_{R}$,
$D_{12}=\langle d_{R}^{\dagger}d_{L}\rangle$,
$D_{21}=\langle d_{L}^{\dagger}d_{R}\rangle$,
and $D_{22}=1-n_{L}$.

By using the non-equilibrium Green's function, the current from the
lead $\alpha\beta$ flowing into the system can be obtained
as \cite{prb66-155308}:
\begin{eqnarray}\label{eq:1h}
J_{\alpha\beta}=-\frac{e}{\pi}\Gamma_{\alpha\beta}\int d\varepsilon f_{\alpha\beta}(\varepsilon)ImG_{\alpha\alpha}^{r}-e\Gamma_{\alpha\beta}\langle d_{\alpha}^{\dagger}d_{\alpha}\rangle
\end{eqnarray}

In the expressions of the current and the coefficients $D_{ij}$,
$\langle d_{\alpha}^{\dagger}d_{\alpha'}\rangle$ is determined
self-consistently. From the relation $\langle
d_{\alpha}^{\dagger}d_{\alpha'}\rangle =-i\int (d\varepsilon/2\pi)
G^<_{\alpha\alpha'}(\varepsilon)$ with the lesser Green's function
$G^<_{\alpha\alpha'}(\varepsilon)$, the self-consistent equations
can exactly be derived \cite{prb66-155308}:
\begin{eqnarray}\label{eq:1i}
&-t_{c}\langle d_{R}^{\dagger}d_{L}\rangle+tc\langle d_{L}^{\dagger}d_{R}\rangle-i\Gamma_{LS}\langle d_{L}^{\dagger}d_{L}\rangle-i\Gamma_{LD}\langle d_{L}^{\dagger}d_{L}\rangle \nonumber \\
=&\int\frac{d\varepsilon}{2\pi}(\Gamma_{LS}f_{LS}+\Gamma_{LD}f_{LD}) (G_{LL}^{r}-G_{LL}^{a}),
\end{eqnarray}
\begin{eqnarray}\label{eq:1j}
&(-\varepsilon_{L}+\varepsilon_{R}-\frac{i}{2}\Gamma_{LS}-\frac{i}{2} \Gamma_{LD}-\frac{i}{2}\Gamma_{RS} \nonumber \\
&-\frac{i}{2}\Gamma_{RD})\langle d_{L}^{\dagger}d_{R}\rangle+t_{c}\langle d_{L}^{\dagger}d_{L}\rangle-t_{c}\langle d_{R}^{\dagger}d_{R}\rangle \nonumber \\
=&\int\frac{d\varepsilon}{2\pi}(\Gamma_{LS}f_{LS}+\Gamma_{LD} f_{LD})G_{RL}^{r}-\int\frac{d\varepsilon}{2\pi}(\Gamma_{RS}f_{RS}+\Gamma_{RD} f_{RD})G_{RL}^{a},
\end{eqnarray}
\begin{eqnarray}\label{eq:1k}
&-t_{c}\langle d_{L}^{\dagger}d_{R}\rangle+tc\langle d_{R}^{\dagger}d_{L}\rangle-i\Gamma_{RS}\langle d_{R}^{\dagger}d_{R}\rangle-i\Gamma_{RD}\langle d_{R}^{\dagger}d_{R}\rangle \nonumber \\
=&\int\frac{d\varepsilon}{2\pi}(\Gamma_{RS}f_{RS}+\Gamma_{RD} f_{RD})(G_{RR}^{r}-G_{RR}^{a}),
\end{eqnarray}
\begin{eqnarray}\label{eq:1l}
&(\varepsilon_{L}-\varepsilon_{R}-\frac{i}{2}\Gamma_{LS}-\frac{i} {2}\Gamma_{LD}-\frac{i}{2}\Gamma_{RS} \nonumber \\
&-\frac{i}{2}\Gamma_{RD})\langle d_{R}^{\dagger}d_{L}\rangle+t_{c}\langle d_{R}^{\dagger}d_{R}\rangle-t_{c}\langle d_{L}^{\dagger}d_{L}\rangle \nonumber \\
=&\int\frac{d\varepsilon}{2\pi}(\Gamma_{RS}f_{RS}+\Gamma_{RD} f_{RD})G_{LR}^{r}-\int\frac{d\varepsilon}{2\pi}(\Gamma_{LS}f_{LS}+\Gamma_{LD} f_{LD})G_{LR}^{a}.
\end{eqnarray}

If we substitute the initial values of $\langle
d_{L}^{\dagger}d_{L}\rangle$, $\langle d_{L}^{\dagger}d_{R}\rangle$,
$\langle d_{R}^{\dagger}d_{L}\rangle$, and $\langle
d_{R}^{\dagger}d_{R}\rangle$ into equations (\ref{eq:1i})-(\ref{eq:1l}),
and solve them self-consistently, we can get the convergent values
of them. Then substituting $\langle d_{L}^{\dagger}d_{L}\rangle$,
$\langle d_{R}^{\dagger}d_{R}\rangle$, and the Green's function of
equation (\ref{eq:1g}) into equation (\ref{eq:1h}), we can get the current.
Besides, the conductance can also be calculated. During the process of
calculations, there is one thing should be emphasized. In general, the lesser
Green's function $G^{<}(\varepsilon)$ can not be solved exactly
for interacting systems. However, in our calculations, we do not
have to solve $G^{<}(\varepsilon)$ itself. When we calculate the
self-consistent equations and the electric current, the quantity
we actually need is $\int d\varepsilon G^{<}(\varepsilon)$ rather
than $G^{<}(\varepsilon)$. Because $\int d\varepsilon G^{<}(\varepsilon)$
can be solve exactly in our model, we need not any approximation
involved in computing $\int d G^{<}(\varepsilon)$ \cite{prb66-155308}.

As we know, the conductance of the two-terminal system is defined as
$G=\frac{dI}{dV}$. While in the DQD system, there are four wires,
i.e., four terminals, and we can define $4\times 4 =16$ conductances
in principle. In the following, we define the conductance as:

\begin{eqnarray}\label{eq:1c}
&G_{\alpha\beta}(V_{\alpha\beta},V_{\alpha\bar{\beta}}, V_{\bar{\alpha}\beta},V_{\bar{\alpha}\bar{\beta}}) \nonumber \\
=&\lim\limits_{V \rightarrow 0}\frac{\left[I_{\alpha\beta}(V_{\alpha\beta}+\frac{V}{2}, V_{\alpha\bar{\beta}}-\frac{V}{2},V_{\bar{\alpha}\beta}, V_{\bar{\alpha}\bar{\beta}}) -I_{\alpha\beta}(V_{\alpha\beta},V_{\alpha\bar{\beta}}, V_{\bar{\alpha}\beta},V_{\bar{\alpha}\bar{\beta}})\right]}{V} ,
\end{eqnarray}
where $\beta$ and $\bar{\beta}$ denote different source-drain leads.
This definition of the conductance is the quantity measured in the
recent experiment \cite{prl110-046604}. Here, we mainly focus on two
different ways of the applied external voltages in the numerical
calculations. One is keeping $V_{LD}=V_{RD}=0$, $V_{LS}=V_{RS}$, and
changing $V_{LS}$ and $V_{RS}$ simultaneously. The other is keeping
$V_{RS}=V_{LD}=V_{RD}=0$, and changing $V_{LS}$ alone. The former
way changes the chemical potentials of both pseudospin up electron
and pseudospin down one simultaneously, which is similar to the
experiments related to the real spin because the chemical potentials
of the spin up electron and the spin down one are difficult to
change separately. Therefore, we could get the
pseudospin-non-resolved results in this way. The latter way changes
the chemical potential of the pseudospin up electron only, thus we
can obtain the pseudospin-resolved results, which is the key point
in \cite{prl110-046604}. We will compare these two ways
carefully under different external conditions in this paper.

\section{Numerical results and analysis}
\label{sec:discussions}

In this section, we at first discuss the case of negligible interdot
tunneling, then generalize our study to the case of finite interdot
tunneling, and at last study the pseudospin flipping current in the
DQD. In our calculations, we have taken
$\Gamma_{LS}=\Gamma_{LD}=\Gamma_{RS}=\Gamma_{RD}=1$ in all cases.

\subsection{\label{sec:dis1}The numerical results without interdot tunneling \label{sec:dis1}}

\begin{figure}
\centering
\includegraphics[width=0.6\textwidth]{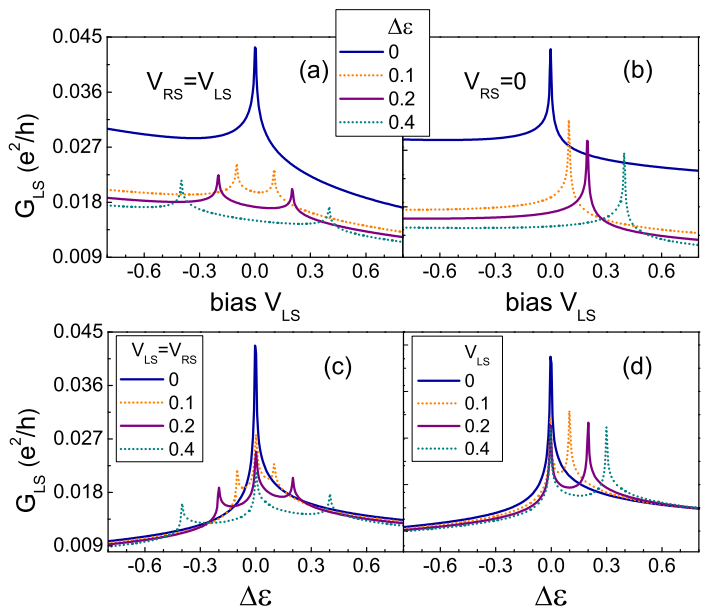}
\caption{(a) and (b) $G_{LS}$ as a function of $V_{LS}$
at different $\Delta\varepsilon$. (c) and (d) $G_{LS}$ as a function of
$\Delta\varepsilon$ at different $V_{LS}$. In (a) and (c), the voltages
$V_{LS}$ and $V_{RS}$ are changed simultaneously, with
$V_{LS}=V_{RS}$ and $V_{LD}=V_{RD}=0$. In (b) and (d), only $V_{LS}$
is changed, with $V_{RS}=V_{LD}=V_{RD}=0$. Other parameters are
$T=0.001$, $\bar{\varepsilon}=-5.0$, and $t_{c}=0$.} \label{fig2}
\end{figure}

In this subsection, we focus on the case without any interdot
tunneling. Before we discuss the conductance of the DQD, there is
one thing to be emphasized. When the interdot tunneling is ignored,
i.e., $t_{c}=0$, there is no pseudospin flipping and we could get
the results of $G_{LS}=G_{LD}$ and $G_{RS}=G_{RD}$. When  finite
interdot tunneling exists, i.e., $t_{c}\neq 0$, all of the four
conductances may not be the same, which is determined by the
structure of the DQD's energy levels and voltages. In addition,
since the characters of the four conductances are similar, we only
analyse $G_{LS}$. Figure \ref{fig1}(b) shows the conductance $G_{LS}$
as functions of $\bar{\varepsilon}$ and $\Delta\varepsilon$, where
$\bar{\varepsilon}=\frac{\varepsilon_{L}+\varepsilon_{R}}{2}$ and
$\Delta\varepsilon=\varepsilon_{L}-\varepsilon_{R}$. The different
color represents different values of the conductance. We can see a
bright peak emerging at $\Delta\varepsilon=0$, which is the
zero-bias Kondo resonant peak.

Next we study the conductance in detail. Figures \ref{fig2}(a) and
\ref{fig2}(b) show $G_{LS}$ as a function of the bias voltage
$V_{LS}$, while figures \ref{fig2}(c) and \ref{fig2}(d) illustrate
$G_{LS}$ as a function of the pseudospin splitting energy
$\Delta\varepsilon$. In figure \ref{fig2}(a), we keep $V_{LS}=V_{RS}$
and $V_{LD}=V_{RD}=0$, implying that the chemical potentials of both
pseudospin up and down electrons are changed simultaneously. When
$\Delta\varepsilon=0$, the Kondo peak emerges at $V_{LS}=0$; when
$\Delta\varepsilon\neq 0$, the Kondo peak splits into two peaks at
$V_{LS}=\pm\Delta\varepsilon$. This phenomenon is similar to the
splitting of the spin Kondo peak of a single QD in the magnetic
field, and $\Delta\varepsilon$ is equivalent to the Zeeman energy
due to the magnetic field. This is the pseudospin-non-resolved Kondo
effect. In figure \ref{fig2}(b), we keep $V_{RS}=V_{LD}=V_{RD}=0$ and
change $V_{LS}$ only. Since the chemical potential of the pseudospin
up electron in the source wire is changed only, there is a single
peak emerging at $V_{LS}=\Delta\varepsilon$. This is the
pseudospin-resolved effect. The results in figures \ref{fig2}(a) and
\ref{fig2}(b) are in good agreement with the recent
experiment \cite{prl110-046604}.

Next we discuss the relation between the conductance $G_{LS}$ and
the pseudospin splitting energy $\Delta\varepsilon$ which is shown
in figures \ref{fig2}(c) and \ref{fig2}(d). In figure \ref{fig2}(c), we
keep $V_{LD}=V_{RD}=0$ and $V_{LS}=V_{RS}$. If $V_{LS}=V_{RS}=0$,
there exists only one Kondo peak which locates at
$\Delta\varepsilon=0$. This is well-known in the spin Kondo system.
While $V_{LS}=V_{RS}\neq 0$, the Kondo peak is divided into three
peaks with their positions locating at $\Delta\varepsilon=0,\pm
V_{LS}$. In figure \ref{fig2}(d), we keep $V_{RS}=V_{LD}=V_{RD}=0$ and
change $V_{LS}$ alone. Different from figure \ref{fig2}(c), only two
peaks are found at $\Delta\varepsilon=0,V_{LS}$ in figure
\ref{fig2}(d) when $V_{LS}\neq 0$, and the original peak at
$\Delta\varepsilon=-V_{LS}$ disappear because the chemical potential
of the pseudospin down electron in the source wire is zero exactly.
It should be noted that since the spin-up and spin-down chemical
potentials in the real spin system are difficult to manipulate
separately, it is not easy to observe these phenomena as shown in
figures \ref{fig2}(c) and \ref{fig2}(d). However, these phenomena are
easy to be observed in the parallel DQD system because it is easy to
manipulate the chemical potentials and the splitting of the
pseudospin degree of freedom.

\begin{figure}
\centering
\includegraphics[width=0.6\textwidth]{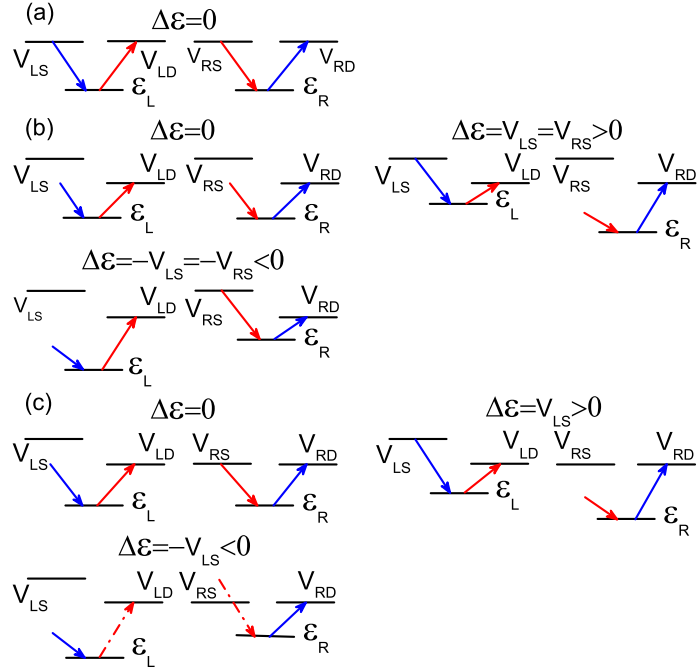}
\caption{Schematic diagram of the electron
cotunneling processes between the DQD and the leads. (a) shows the case
when $V_{LS}=V_{RS}=V_{LD}=V_{RD}=0$ and $\Delta\varepsilon=0$. In (a),
an electron tunnels from the right QD into the lead $RD$ and another
electron tunnels from the lead $LS$ into the left QD, which is shown by
the blue arrows. The red arrows show two similar tunneling events:
an electron tunnels from the left QD to the lead $LD$ and another one
tunnels from the lead $RS$ into the right QD. By combining these four events,
the electrons can pass through both QDs. (b) shows the similar
cotunneling processes when $V_{LD}=V_{RD}=0$ and $V_{LS}=V_{RS}\neq
0$. (c) shows the similar cotunneling processes when
$V_{RS}=V_{LD}=V_{RD}=0$ and $V_{LS}\neq 0$. Both (b) and (c)
illustrate three different cases: $\Delta\varepsilon=0$,
$\Delta\varepsilon>0$, and $\Delta\varepsilon<0$. The dash-dotted lines in (c)
indicate that the tunneling events are forbidden.} \label{fig3}
\end{figure}

The Kondo peaks in figure \ref{fig2} can be understood by the
cotunneling processes shown in figure \ref{fig3}.
It should be pointed out that the Kondo effect can be captured by
the fourth or higher-order perturbation processes with respect to
the tunneling between dot and leads.
As we can see from figure \ref{fig3},
when the electric state in the parallel DQDs returns to its original
state, it has experienced four tunneling processes (shown by two
red lines and two blue lines).
These four tunneling processes make
up two cotunneling processes, and each cotunneling process is a second
order perturbation process. Notice that Only the combination of two 
cotunneling processes can lead to the Kondo effect. The similar 
explanation, which interprets the Kondo effect by cotunneling
precesses, has been used in many previous papers \cite{prb66-155308,
prb88-235427,prb68-155323,pe42-2446,jmmm272-1676}. 
Figure \ref{fig3}(a) plots a cotunneling process which leads to the 
main Kondo resonance when $V_{LS}=V_{LD}=V_{RS}=V_{RD}=0$ and $\Delta\varepsilon=0$ 
(blue lines in figures \ref{fig2}(c) and \ref{fig2}(d)). The blue and 
red arrows illustrate the correlative tunneling events, respectively. 
To be specific, we first consider an electron in the right QD. This
electron can tunnel from the right QD into the lead $RD$. Then,
another electron in the lead $LS$ with the energy $V_{RD}$ can
tunnel into the left QD. These two tunneling events are shown by the
blue arrows. After that, the left QD is occupied and the right QD is
empty, where the system energy is the same as that in the beginning
state. The red arrows show another two similar tunneling events,
where an electron in the left QD tunnels into the lead $LD$ and then
another electron in the lead $RS$ tunnels into the right QD. With
the above four tunneling events, although the system recovers to the
beginning state, the electrons travel from the left (right) source
lead through the left (right) QD to the left (right) drain lead.
When many of these cotunneling processes take coherent superposition
at low temperature, a Kondo resonance will appear. This leads to the
main Kondo peak at $\Delta\varepsilon=0$ in figure \ref{fig2}(c) for
$V_{LS}=V_{RS}=0$ and in figure \ref{fig2}(d) for $V_{LS}=0$. Figure
\ref{fig3}(b) explains the emergence of three peaks when
$V_{LS}=V_{RS}\neq 0$ in figure \ref{fig2}(c). No matter
$\Delta\varepsilon=0$, $\Delta\varepsilon>0$, or
$\Delta\varepsilon<0$, the electrons can travel through both QDs
because of the cotunneling processes shown in figure \ref{fig3}(b),
and thus three peaks appear in figure \ref{fig2}(c). On the other
hand, it should be pointed out that the energy is conserved in the
cotunneling processes. Therefore, for $\Delta\varepsilon=0$ in figure
\ref{fig3}(b), when an electron in the right QD tunnels into the
lead $RD$, another electron in the lead $LS$ with the energy
$V_{RD}$ can tunnel into the left QD. For $\Delta\varepsilon>0$
($\Delta\varepsilon<0$), the condition of
$V_{LS}-\varepsilon_L=V_{RD}-\varepsilon_R$
($V_{LD}-\varepsilon_L=V_{RS}-\varepsilon_R$) should be preserved
due to the energy conservation in the cotunneling processes. Since
we keep $V_{LD}=V_{RD}=0$ and $V_{LS}=V_{RS}$, the Kondo peaks can
emerge at $\Delta\varepsilon\equiv\varepsilon_L-\varepsilon_R=\pm
V_{LS}=\pm V_{RS}$ (see figure \ref{fig2}(c)). Figure \ref{fig3}(c)
explains the emergence of two peaks when $V_{LS}\neq 0$ in figure
\ref{fig2}(d). For $\Delta\varepsilon=0$ and $\Delta\varepsilon>0$,
the electrons can pass through both QDs. However, for
$\Delta\varepsilon<0$, the energy obtained from the electron jumping
from the right source lead $RS$ to the right QD cannot support the
tunneling event from the left QD to the drain lead $LD$ (shown by
the red dash-dotted lines). Therefore, the electrons cannot travel
through the DQD for $\Delta\varepsilon<0$. As a result, no Kondo
peak appears at $\Delta\varepsilon =-V_{LS}$ and there are only two
Kondo peaks at $\Delta\varepsilon=0$ and $\Delta\varepsilon=V_{LS}$
in figure \ref{fig2}(d).

\begin{figure}
\centering
\includegraphics[width=0.6\textwidth]{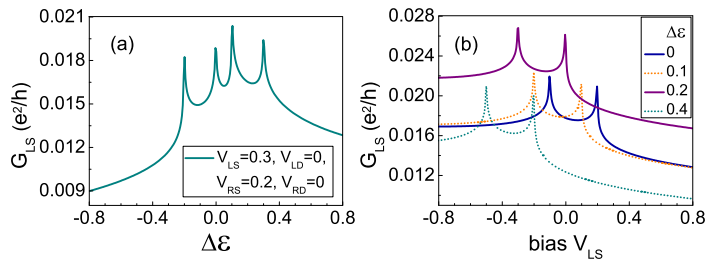}
\caption{(a) Conductance $G_{LS}$ as a function of
the pseudospin splitting energy $\Delta\varepsilon$ with $V_{LS}=0.3$, $V_{RS}=0.2$, and $V_{LD}=V_{RD}=0$. (b) $G_{LS}$ versus the voltage
$V_{LS}$ at different $\Delta\varepsilon$ with $V_{RS}=0.2$, $V_{LD}=0.0$, and $V_{RD}=-0.1$. The remaining parameters are $T=0.001$, $\bar{\varepsilon}=-5.0$, and $t_{c}=0$.} \label{fig4}
\end{figure}

In general, if the four lead voltages $V_{LS}$, $V_{LD}$, $V_{RS}$,
and $V_{RD}$ do not equal to each other, there are four Kondo peaks
with their positions at $\Delta \varepsilon=V_{LS}-V_{RS}$, $\Delta
\varepsilon=V_{LS}-V_{RD}$, $\Delta \varepsilon=V_{LD}-V_{RS}$, and
$\Delta \varepsilon=V_{LD}-V_{RD}$, respectively. It should be noted
that although it can have four Kondo peaks in the curve of the
conductance as a function of the pseudospin splitting $\Delta
\varepsilon$, there are at most two Kondo peaks in the curve of the
conductance versus the voltage, e.g., $G_{LS}$ versus $V_{LS}$. When
some of the four lead voltages have identical value, some Kondo
peaks will overlap and then the number of the peaks can be reduced,
as shown in figure \ref{fig2}. Figure \ref{fig4}(a) displays $G_{LS}$ versus
$\Delta\varepsilon$ with $V_{LS}=0.3$, $V_{RS}=0.2$, and
$V_{LD}=V_{RD}=0$, in which four Kondo peaks clearly exhibit. Figure
\ref{fig4}(b) shows $G_{LS}$ versus the voltage $V_{LS}$ by fixing
$V_{RS}=0.2$, $V_{LD}=0$, and $V_{RD}=-0.1$ with different
pseudospin splitting energy $\Delta \varepsilon$. Here, two Kondo
peaks emerge. It is worth mentioning that the conductance $G_{LS}$
at $\Delta \varepsilon =-0.2$ is obviously larger than the other
cases. This is due to the fact that when $\Delta \varepsilon =-0.2$,
$\Delta \varepsilon=V_{LD}-V_{RS}$ keeps, regardless of the voltage
$V_{LS}$. This means that the Kondo resonance occurs always, so a
very large conductance $G_{LS}$ could be observed at low
temperature.

\subsection{\label{sec:dis2}The effect of the interdot tunneling}

\begin{figure}
\centering
\includegraphics[width=0.6\textwidth]{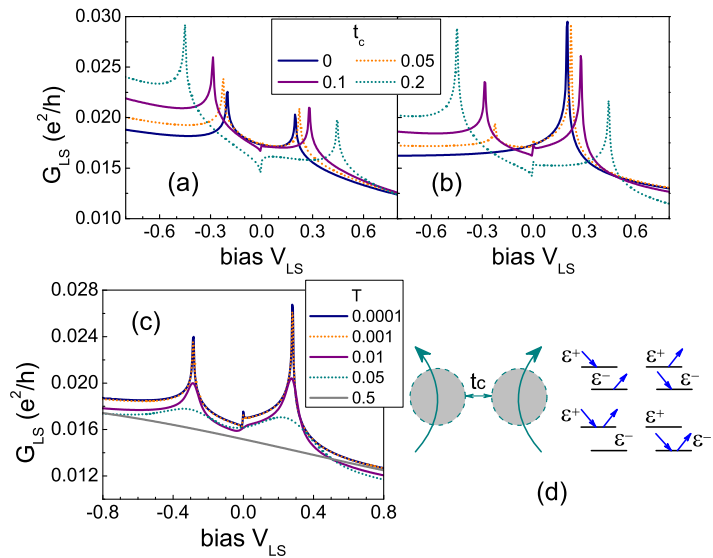}
\caption{(a) and (b) Conductance $G_{LS}$ as a
function of the voltage $V_{LS}$ at different $t_{c}$. In (a), $V_{LS}$ and $V_{RS}$ are changed simultaneously, with $V_{LD}=V_{RD}=0$.
In (b), only $V_{LS}$ is changed, with $V_{RS}=V_{LD}=V_{RD}=0$.
The temperature is $T=0.001$. (c) $G_{LS}$ as a
function of $V_{LS}$ at different temperature $T$. In (c), only
$V_{LS}$ is changed, with $V_{RS}=V_{LD}=V_{RD}=0$ and
$t_{c}=0.1$. The remaining parameter is $\bar{\varepsilon}=-5.0$. (d) Schematic diagram of the four cotunneling processes between the molecular states and the leads.} \label{fig5}
\end{figure}

When the interdot tunneling coupling $t_c$ is considered, we can
generalize the experimental results of \cite{prl110-046604}.
Before the discussion of the
conductance, let us first analyse the cotunneling processes at $t_c
\not=0$. Figure \ref{fig5}(d) shows the change of the energy level of
the DQD in the presence of $t_{c}$. When $t_{c}\not=0$, the energy
levels in the left and right QDs will hybridize into the molecular
states. That is, $\varepsilon_{L}$ and $\varepsilon_{R}$ can be
recombined into $\varepsilon^{\pm}=\frac{(\varepsilon_{L}+\varepsilon_{R})}
{2}\pm\frac{\Delta E}{2}$ which expands to the entire device at
$\Delta\varepsilon=0$ \cite{prb66-155308,nature395-873,prb61-12643},
where $\Delta E=\sqrt{\Delta\varepsilon^{2}+4t_{c}^{2}}$. Then,
there will be four kinds of cotunneling processes in the DQD (see
figure \ref{fig5}(d)). (1) The electron originally occupying
$\varepsilon^{-}$ tunnels to the lead $RD$ ($LD$), and another
electron at $V_{RD}+\Delta E$ ($V_{LD}+\Delta E$) in the lead $LS$
($RS$) tunnels to $\varepsilon^{+}$. (2) The electron at the state
$\varepsilon^{+}$ tunnels to the lead $LD$ ($RD$), and another
electron at $V_{LD}-\Delta E$ ($V_{RD}-\Delta E$) in the lead $RS$
($LS$) tunnels to $\varepsilon^{-}$. (3) The electron at the state
$\varepsilon^{+}$ tunnels to the drain lead $LD$ ($RD$), and another
electron at $V_{LD}$ ($V_{RD}$) in the source lead $LS$ ($RS$)
tunnels to $\varepsilon^{+}$. (4) The electron at $\varepsilon^{-}$
tunnels to the lead $LD$ ($RD$), and another electron at $V_{LD}$
($V_{RD}$) in the lead $LS$ ($RS$) tunnels to $\varepsilon^{-}$.
Here, although the cotunneling processes may be similar to that
discussed in \cite{prb66-155308}, the conductance is
totally different. In \cite{prb66-155308}, the system
is a serial DQD. When $t_{c}=0$, since there is no transport
coupling between the two QDs, $I$ and $\frac{dI}{dV}$ are zero
exactly. In the present system, because each QD is connected to its
own source and drain leads, $I$ and $\frac{dI}{dV}$ are nonzero, no
matter $t_{c}=0$ or $t_{c}\neq 0$.

Figure \ref{fig5}(a) shows the conductance $G_{LS}$ as a function of
the voltage $V_{LS}$ by changing $V_{LS}$ and $V_{RS}$
simultaneously, i.e., $V_{LS}=V_{RS}$. For $t_{c}=0$, the Kondo
peaks locate at $V_{LS}=\pm\Delta\varepsilon$. When $t_{c}$ is
increased, the two Kondo peaks move to $V_{LS}=\pm\Delta E$. Thus,
they could emerge in larger $|V_{LS}|$ with increasing $t_{c}$.
These two peaks correspond to the first and second kind of the
cotunneling processes as discussed in the above paragraph. In
addition, another small Kondo peak and dip emerge at $V_{LS}=0$,
which is attributed to the third and fourth kind of the cotunneling
processes. Notice that in the third and fourth kind of the
cotunneling processes, the original and final electrons are at the
same molecular state. Thus, the Kondo peak and the dip is always
fixed around $V_{LS}=0$. Figure \ref{fig5}(b) shows $G_{LS}$ as a
function of $V_{LS}$ when only $V_{LS}$ is changed and
$V_{LD}=V_{RS}=V_{RD}=0$. At $t_{c}=0$, there is only one Kondo peak
at $V_{LS}=\Delta\varepsilon$, which is the pseudospin-resolved
Kondo peak observed in the experiment of \cite{prl110-046604}.
However, when $t_{c}$ is increased,
this peak moves to $V_{LS}=\Delta E$. Besides, the Kondo peak at
$V_{LS}=-\Delta E$ also emerges, and its height becomes higher and
higher. The reason is that at $t_c\not=0$, the electron at the
molecular state $\varepsilon^-$ ($\varepsilon^+$) can tunnel to both
left and right drain leads, and the electron in the left and right
source leads can tunnel to the molecular state $\varepsilon^-$
($\varepsilon^+$). This is different from $t_c=0$, in which the
electron at the level $\varepsilon_L$ ($\varepsilon_R$) can only
tunnel to one drain lead $LD$ ($RD$). Additionally, a small peak and
a small dip emerge around $V_{LS}=0$, because of the third and
fourth kind of the cotunneling processes. In figure \ref{fig5}(c), we
show the dependence of $G_{LS}$ on temperature $T$. It can be
clearly seen that with increasing $T$, the height of the Kondo peak
becomes lower and lower. At $T=0.5$, all of the Kondo peaks
disappear.

\begin{figure}
\centering
\includegraphics[width=0.6\textwidth]{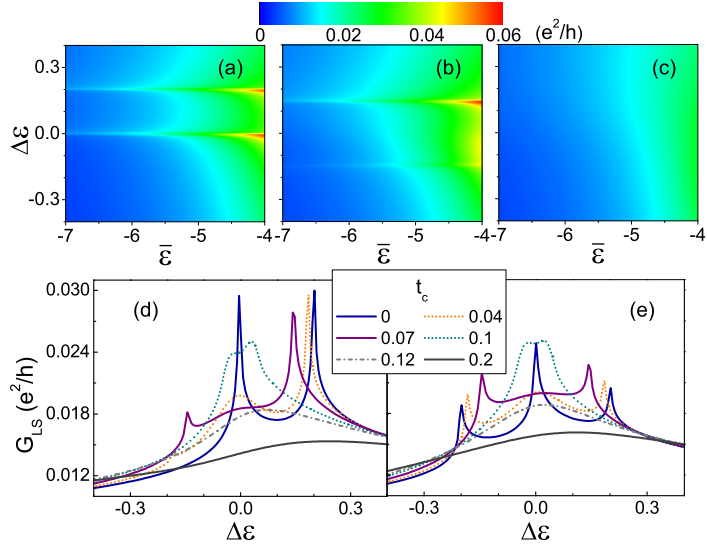}
\caption{(a)-(c) Conductance $G_{LS}$ as functions
of $\bar{\varepsilon}$ and $\Delta\varepsilon$. The source and drain
voltages are set to $V_{LS}=0.2$ and $V_{RS}=V_{LD}=V_{RD}=0$. $t_{c}$ is taken as $0$, $0.07$, and $0.2$ in (a), (b), and (c), respectively. (d) and (e)
show $G_{LS}$ as a function of $\Delta\varepsilon$ at different
$t_{c}$ with $\bar{\varepsilon}=-5.0$. In (d), $V_{LS}=0.2$ and $V_{RS}=V_{LD}=V_{RD}=0$; in (e), $V_{LS}=V_{RS}=0.2$ and $V_{LD}=V_{RD}=0$. The temperature is $T=0.001$.} \label{fig6}
\end{figure}

Next, we investigate the conductance $G_{LS}$ as a function of the
pseudospin splitting energy $\Delta\varepsilon$ at different
$t_{c}$. In figure \ref{fig6}(d), the voltages are set to $V_{LS}=0.2$
and $V_{RS}=V_{LD}=V_{RD}=0$. At $t_{c}=0$, there are two Kondo
peaks at $\Delta\varepsilon=0$ and $\Delta\varepsilon=V_{LS}$. With
increasing $t_{c}$, the original peak at $\Delta\varepsilon=V_{LS}$
moves toward $\Delta\varepsilon=0$ and the height is decreased,
because this Kondo peak now locates at $\sqrt{\Delta \varepsilon^2
+4t_c^2}=V_{LS}$. The other Kondo peak emerges at the symmetric
place of the other side of $\Delta\varepsilon$. In addition, the
peak at $\Delta\varepsilon=0$ broadens and the height is declined.
If $t_{c}$ is gradually increased, the height of the peak at
$\Delta\varepsilon=0$ is decreased. At the same time, the two peaks
at the opposite sides of $\Delta\varepsilon$ move toward
$\Delta\varepsilon=0$, and eventually mix together at
$\Delta\varepsilon=0$. Thus, there is only one broadening peak
around $\Delta\varepsilon=0$. Then, by further increasing $t_{c}$,
the height of this broadening peak decreases until this peak
vanishes. This is attributed to the fact that the two quantum dots
become a whole when $t_{c}$ is considerably large. The degeneracy of
the pseudospin does not exit, so does the Kondo effect. Figures
\ref{fig6}(a)-\ref{fig6}(c) are the two-dimensional plot of the
conductance $G_{LS}$ versus $\Delta \varepsilon$ and
$\bar{\varepsilon}$ with $t_{c}=0$, $0.07$, and $0.2$, respectively.
The change of the color in figures \ref{fig6}(a)-\ref{fig6}(c) clearly
shows the process discussed above. As a comparison, figure
\ref{fig6}(e) shows $G_{LS}$ as a function of the pseudospin
splitting energy $\Delta\varepsilon$ when $V_{LS}=V_{RS}=0.2$ and
$V_{LD}=V_{RD}=0$. It is clear that at $t_{c}=0$, except for the
peak at $\Delta\varepsilon=0$, there are two Kondo peaks at both
sides of $\Delta\varepsilon$. When $t_{c}$ is increased, the peak at
$\Delta\varepsilon=0$ becomes lower and boarder; and the peaks at
both sides move toward $\Delta\varepsilon=0$, and eventually mix
together. If $t_{c}$ is gradually increased, the last peak becomes
lower till it disappears.

\subsection{\label{sec:dis3} Pseudospin transport and pseudospin flipping current}

\begin{figure}
\centering
\includegraphics[width=0.6\textwidth]{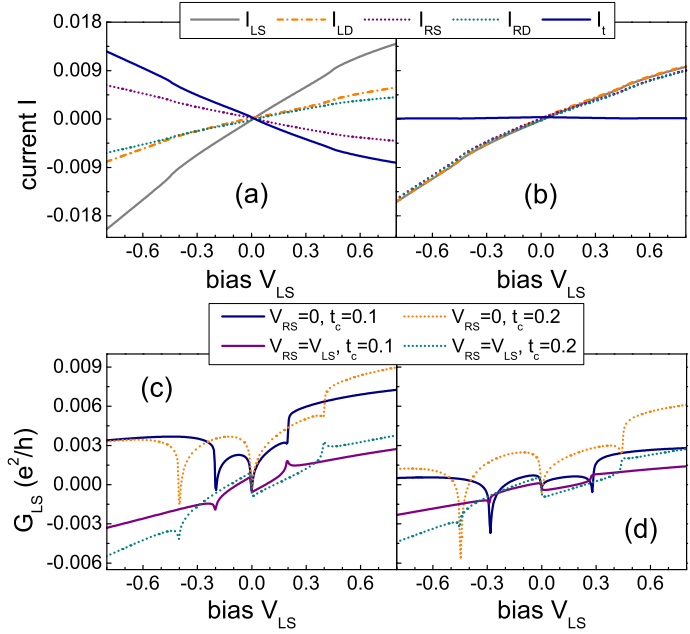}
\caption{(a) and (b) show the current in the leads
$LS$, $LD$, $RS$, and $RD$, and the pseudospin flipping current
$I_{t}$ as a function of $V_{LS}$ with the pseudospin splitting
energy $\Delta\varepsilon=0.2$ and $t_{c}=0.2$. In (a), only
$V_{LS}$ is changed; in (b), both $V_{LS}$ and $V_{RS}$ are changed with $V_{LS}=V_{RS}$. (c) and (d) show the pseudospin flipping conductance $G_{t}$ as a function of $V_{LS}$ with $\Delta\varepsilon=0$ and $\Delta\varepsilon=0.2$, respectively. All the unchanged source and drain voltages are set to zero, the temperature keeps $T=0.001$, and $\bar{\varepsilon}=-5.0$.} \label{fig7}
\end{figure}

In this subsection, we discuss the pseudospin transport in the DQD
system. As we know, the direction of the real spin can be changed in
the electron transport process. As a result, a steady spin current
cannot be held easily. On the other hand, the measurement of the
spin current is also difficult. Thus, it limits the development of
the research field on the spin transport. The orbital Kondo effect,
which is a pseudospin Kondo effect, can be regarded as the
counterpart of the spin Kondo effect. The Kondo effect, whose
emergence is originally related to the spin degree of freedom, can
also be realized in the system with the orbital degree of freedom.
This indicates that we may use a system, including the orbital
degree of freedom, to study the physical properties which are
difficult to be observed with the spin degree of freedom. In the DQD
system, the current flow in the leads $LS$, $RS$, $LD$, and $RD$ is
easy to measure, which means that the pseudospin current is easy to
measure. Furthermore, the pseudospin flipping only happens in the
QDs and its flipping strength is controllable and tunable. When the
current flows in the leads, it cannot tunnel from the left side
($LS$ and $LD$) to the right side ($RS$ and $RD$), which indicates
that the pseudospin current is conserved in the leads. Thus, it is
possible and convenient to use the orbital degree of freedom to
study the properties related to the spin degree of freedom.

It should be pointed out that, when $t_{c}\neq 0$, the currents in
the leads $LS$, $LD$, $RS$, and $RD$ may not be equal to each other,
but they still satisfy the relation $I_{LS}+I_{RS}=I_{LD}+I_{RD}$
due to the electric current conservation. Here, we define that the
positive direction of the current is flowing into the DQD for the
source leads and is going out from the DQD for the drain leads.
Besides, we introduce the pseudospin flipping current $I_t$, which
describes the current from the right QD to the left one. The
relation between $I_t$ and the four wire currents are
$I_{LS}+I_{t}=I_{LD}$ and $I_{RS}-I_{t}=I_{RD}$. Thus, $I_t$ can be
expressed as:
\begin{eqnarray}
I_{t}=\frac{(I_{LD}-I_{RD})-(I_{LS}-I_{RS})}{2} =\frac{I^{spin}_D-I^{spin}_S}{2}, \nonumber
\end{eqnarray}
where $I^{spin}_{S/D} \equiv I_{LS/D}-I_{RS/D}$ is the pseudospin
current in the source/drain lead. Figures \ref{fig7}(a) and \ref{fig7}(b)
illustrate the pseudospin flipping current $I_{t}$ as a function of the
voltage $V_{LS}$. In figure \ref{fig7}(a) only $V_{LS}$ is changed, and
in figure \ref{fig7}(b) both $V_{LS}$ and $V_{RS}$ are changed with
$V_{LS}=V_{RS}$. It is clear that when only $V_{LS}$ is
changed, the pseudospin flipping current $I_{t}$ is considerable as
compared with the current in the four leads, because the
pseudospin-up chemical potential $eV_{LS}$ is not equal to the
pseudospin-down one $eV_{RS}$, i.e., there exists a pseudospin bias
$V^{spin}_S= V_{LS}-V_{RS}$. On the other hand, when both $V_{LS}$
and $V_{RS}$ are changed, the pseudospin flipping current
$I_{t}$ is negligible, because the pseudospin bias
$V^{spin}_{S/D}=V_{LS/D}-V_{RS/D}$ is zero. These calculations
demonstrate that if we deal with the pseudospin-resolved transport
spectroscopy in the DQD system, a steady pseudospin current can be
induced. The magnitude of this pseudospin current is not small, and
in particular it is easy to be controlled and measured.

At last, in order to see the characteristics of the pseudospin
flipping in the DQD more clearly, we calculate the flipping
conductance which is defined as
\begin{eqnarray}\label{eq:1d}
&G_{t}(V_{LS},V_{LD},V_{RS},V_{RD}) \nonumber \\
=&\lim\limits_{V \rightarrow 0}[I_{t}(V_{LS}+\frac{V}{2},V_{LD}-\frac{V}{2},V_{RS},V_{RD})-I_{t}(V_{LS},V_{LD},V_{RS},V_{RD})]/V .
\end{eqnarray}
Notice that in the above definition, only the left source and drain
voltages are changed by $\pm V/2$. Figures \ref{fig7}(c) and
\ref{fig7}(d) show $G_{t}$ as a function of $V_{LS}$ with
$\Delta\varepsilon=0$ and $0.2$, respectively. The results exhibit
the following features: (1) no matter whether $V_{RS}$ is changed
with $V_{LS}$ or not, the Kondo peaks and dips of $G_{t}$ emerge at
$V_{LS}=0$ and $V_{LS}=\pm\Delta E$; (2) with increasing of $t_{c}$,
$G_{t}$ is enhanced in usual; (3) the dips are much sharper when
$V_{LS}$ is changed only, which also indicates that we can focus on
the pseudospin-resolved transport spectroscopy when we study the
pseudospin flipping current in the parallel DQD systems.

\section{Conclusion}
\label{sec:conclusions}

In this paper, we investigate the orbital Kondo effect in a parallel
double quantum dot. When the interdot tunneling coupling $t_{c}$ is
zero, we explain the pseudospin-resolved results observed in the
recent experiment \cite{prl110-046604}. We find that there exist
three Kondo peaks and two Kondo peaks in the curve of the
conductance versus the pseudospin splitting energy for the
pseudospin-non-resolved case and the pseudospin-resolved case,
respectively. When the interdot coupling $t_{c}$ is nonzero, the
levels in the separated quantum dots can hybridize into the
molecular levels, and new Kondo peaks emerge. In addition, the
pseudospin flipping current and the conductance are also studied,
and both of them show the Kondo peaks and dips. We point out that
the present pseudospin system has many advantages in comparison with
the real spin system. In the pseudospin system, the chemical
potential of each pseudospin component, the pseudospin splitting
energy, and the coupling strength can be well controlled and tuned.
Besides, the pseudospin current is conserved in the source and drain
leads, and the pseudospin-up and pseudospin-down currents can
individually be measured. Therefore, we believe that these results
could be observed in the present technology.

\section*{Acknowledgements}

This work was financially supported by NBRP of China (2012CB921303)
and NSF-China under Grants No. 11274364.

\end{document}